\documentclass[12pt]{iopart}
\usepackage{graphics}

\newcommand  {\Rbar} {{\mbox{\rm$\mbox{I}\!\mbox{R}$}}}

\newcommand{\Lie}[0]{{\cal L}\, }

\newcommand{\be}{\begin{equation}}
\newcommand{\ee}{\end{equation}}
\newcommand{\bea}{\begin{eqnarray}}
\newcommand{\eea}{\end{eqnarray}}
\newcommand{\nn}{\nonumber}

\newcommand{\ssg}{{\sqrt{\sigma}}}

\newcommand{\cH}{\mathcal{H}}

\newcommand{\part}{T}

\newcommand{\tid}{\bar}

\begin{document}

\title{Canonical Phase Space Formulation of Quasi-local General Relativity}
\author {Ivan Booth\dag and Stephen Fairhurst\ddag}

\address{\dag Department of Mathematics and Statistics,
Memorial University of Newfoundland \\
St. John's, Newfoundland, A1C 5S7, Canada}

\address{\ddag
Theoretical Physics Institute, Department of Physics, University
of Alberta \\  Edmonton, Alberta, T6G 2J1, Canada}

\begin{abstract}
We construct a Hamiltonian formulation of quasi-local general
relativity using an extended phase space that includes boundary
coordinates as configuration variables. This allows us to use
Hamiltonian methods to derive an expression for the energy of a
non-isolated region of space-time that interacts with its
neighbourhood. This expression is found to be very similar to the
Brown-York quasi-local energy that was originally derived by
Hamilton-Jacobi methods. We examine the connection between the two
formalisms and find that when the boundary conditions for the two are
harmonized, the resulting quasi-local energies are identical.
\end{abstract}

\maketitle

\section{Introduction}

General relativity is a fully covariant theory of gravity and as such
does not privilege any particular flow of time. This is one of its
great virtues but it is also a problem if one wishes to study it using
traditional methods that manifestly depend on a notion of time. For
example, a flow of time must be defined before one can 
do a standard Hamiltonian phase space analysis of a field theory. Thus, 
if we want to apply such
methods to Einstein's gravity, space-time must be artificially broken
into space and time. We do this by foliating space-time into a set of
space-like three-surfaces (``instants of time'') $\Sigma_t$, and
defining a ``time-flow'' vector field $T^a$ that evolves these
surfaces into each other.

Having done this, one can reformulate general relativity in terms of
phase space, symplectic structures, and Hamiltonians. In the most
common approach, the space-like three-metric $h_{ab}$ is taken as the
configuration variable while its conjugate momentum, $P^{ab}$ is
closely related to the extrinsic curvature of the three-surface in
four-dimensional space-time (see for example \cite{adm,rt,wald}). 
Traditionally, space-times studied in this way were
either boundary-free or taken to have a boundary at spatial or null
infinity (with appropriate conditions imposed so that the space-time is
asymptotically flat).  More recently, people have become interested in
studying general relativity over finite regions of space-time, which
are often referred to as quasi-local regions. Then, boundaries and
boundary conditions at locations other than infinity must be
considered. In \cite{quasiphase,kij}, the assorted boundary conditions
that give rise to a phase space with a well-defined symplectic
structure were studied. In particular, it is well known that this can
be done if one fixes the intrinsic metric $\gamma_{ab}$ of the
boundary to be constant and not affected by variations. That is, one
fixes the intrinsic geometry at each point of the boundary manifold.

However, some aspects of this fix are not entirely satisfactory. For
example, if the fixed boundary metric is not axially symmetric, then
a rotation along a vector field $\phi$ which is tangent to the
boundary at the boundary \textit{is not} an allowed variation in the phase 
space under consideration, because $\delta_{\phi} \gamma_{ab} = \Lie_{\phi}
\gamma_{ab} \neq 0$ and so the variation does not preserve the
boundary condition.  Thus, there is no Hamiltonian generating this
motion in phase space and consequently no notion of the angular
momentum associated to such a rotation.  Similarly, time translations
are not permitted unless the boundary metric is invariant in time,
i.e. $\delta_T \gamma_{ab} = \Lie_T \gamma_{ab} = 0$.  These
evolutions cannot be generated by Hamiltonian functionals unless 
$\phi^a$ and $T^a$ are Killing vector fields of $\gamma_{ab}$.

A little thought shows that this is really not so surprising.  In both
of the examples discussed above, the ``conserved'' quantity
corresponding to the listed translation is not conserved.  Hence, we
should not expect to be able to obtain it from a standard Hamiltonian
treatment, which by its very nature applies to situations where
Hamiltonians are conserved. For example, if a Hamiltonian exists that
generates time translations, then Hamilton's equations are
\begin{equation}
   \label{hameq} \delta H_{T} = \Omega(\delta_T, \delta)
\end{equation}
where the symplectic structure $\Omega$ is antisymmetric in the two
variations. Hence, it follows immediately that $\delta_{T} H_{T} =
0$. Similarly, the Hamiltonian generating a particular rotation is
conserved under that same rotation.

Despite the arguments given above, Brown and York have introduced
notions of the energy and angular momentum associated with a bounded
region of space-time (if the intrinsic boundary metric is fixed) by
performing a Hamilton--Jacobi analysis \cite{BY}. Furthermore, the
expressions they have obtained have proved useful in a very large
number of applications (as a representative sample see \cite{BYapp}).
It would be very surprising if these expressions could not be derived
from a phase space treatment of general relativity over a manifold
with boundary.

The purpose of this paper is to show that it is indeed possible to
obtain these expressions for energy and angular momentum using a
careful phase space analysis of general relativity on manifolds with
boundaries.  Here, we begin with the standard phase space formulation
of general relativity in terms of ADM variables (see, for example,
\cite{ashbook} for details).  The key idea is then to import extended
phase space techniques from classical mechanics which are designed to
deal with situations where ``conserved'' quantities are not conserved.
The basic idea is to enlarge the phase space under consideration by
including quantities such as the time coordinate as configuration
variables (see \cite{arnold} for a standard reference).  It is then
possible to construct a conserved Hamiltonian generating time
translations in the extended phase space.  However, the energy of the
system is no longer the value of this Hamiltonian, but is instead the
value of the momentum which is canonically conjugate to time.

In addition to extending the phase space however, it is also necessary
to weaken the boundary conditions so that time translations and
rotations \textit{are} allowed variations.  For example, in the
Brown-York treatment the boundary three-metric $\gamma_{ab}$ is
completely fixed on a three-manifold $B$ that is a boundary of the
space-time manifold $M$.  Therefore,
\be
\delta \gamma_{ab} = 0
\ee
for all variations. That is, the metric is fixed with respect to the
manifold.  Under our looser treatment, we allow the variations to act
as diffeomorphisms that move our boundary fields around $B$. Thus,
$\delta$ acts on the ``fixed'' fields as an infinitesimal
diffeomorphism and so
\be 
\delta \gamma_{ab} = \Lie_{\delta X} \gamma_{ab}, 
\ee
for some vector field $\delta X^a$. Picking $\delta X^a$
appropriately, we obtain translations and rotations as variations.

The generalization of the Hamiltonian formulation for a finite region
of space-time is, to our knowledge, presented here for the first time.
However, similar ideas have previously been discussed at spatial
infinity.  Notably, in their early work on boundary terms for the
gravitational Hamiltonian, Regge and Teitelboim \cite{rt} implemented
a similar programme for the boundary at spatial infinity in
asymptotically flat space-times. Specifically they included variables
that located the asymptotic position of the boundary as well as their
conjugate momenta as canonical variables and showed that the resulting
Hamiltonian was covariant with respect to the asymptotic Poincar\'{e}
group. Kucha\v{r} has also studied parameterizations at infinity.  In
his considerations of Hamiltonian formulations for spherically
symmetric black holes \cite{kuchar}, he included the Killing time as a
canonical coordinate and found that the mass was conjugate to its
radial rate of change.  Finally, Kijowski \cite{kij} has also
considered general relativity in a manifold with boundary.  He begins
with a novel and non-standard approach to symplectic geometry and the
Hamiltonian formulation \cite{kt}.  Despite this different starting
point, he obtains energy expressions for the boundary which are
analogous to the Brown--York.  However, he only explicitly considers
the case where the evolution is a symmetry of the boundary.  Thus, our
work may be considered a generalization to arbitrary boundary
geometries.  It is likely that our method could also be incorporated
into this alternative formalism.

The algebra for the quasi-local gravitational case is quite formidable,
and so as an introduction, we begin in section \ref{s2} with a simple
example of the extended phase space formalism, namely a time dependent
harmonic oscillator.  With this experience in hand, in section
\ref{s3} we apply similar techniques to the problem of interest:
general relativity in a space-time manifold with a boundary.  In
particular, we obtain expressions for the energy and angular momentum 
associated to the region of space-time, which may be calculated using 
just the values of the fields at that boundary.  Section
\ref{s4} provides a comparison between our results and those of Brown
and York.  We end with a discussion of the results and possible
extensions and applications of this work.  Several key technical
results are collected in the appendices.

\section{An Introductory Example}
\label{s2}

In this section, we shall consider a simple example which will
demonstrate how a phase space can be extended to allow the description
of systems in which the energy is not constant. By starting with a
time-dependent harmonic oscillator we will capture many of the central
ideas of the construction without the extra complications that arise
in gravity. The key idea will be to extend the phase space by
including the time coordinate $t$ as a configuration variable as well
as its conjugate momentum $p_t$. Then if we know how the variables
evolve in time, we can manipulate the symplectic structure so as to
find a Hamiltonian function which generates that evolution
on-shell. Equivalently, we solve the Hamiltonian equations of motion
for the given evolution to find the corresponding Hamiltonian.

With the standard phase space treatment, the energy of the system is
the on-shell value of this generator of time translations (which is,
of course, constant). In the extended phase space treatment, the
energy associated with a time translation is equal to the negative of
the value of the momentum conjugate to $t$. A slight complication is
that, in general, the Hamiltonian is not unique and as a result of
this ambiguity the energy conjugate to the evolution is not unique
either. Indeed a significant degree of freedom remains in its
definition.  We illustrate and elaborate on these issues in the
following example.

\subsection{Simple harmonic oscillator}
\label{s2.1}

Consider the harmonic oscillator with mass $m$ and spring constant
$k$.  The canonical phase space of this system is parameterized by the
position coordinate $q$ and its conjugate momentum $p$.  The
symplectic structure is simply given by
\be
\Omega(\delta_1, \delta_2) = 
(\delta_1 q)(\delta_2 p) - (\delta_2 q)(\delta_1 p)\, .
\ee
In this and all future expressions, one should keep in mind that the
$\delta$s are vectors in the phase space of all possible 
configurations of the system. Their conventional interpretation as
infinitesimal variations of system configurations arises from considering the
one-parameter families of phase space diffeomorphisms that are generated by 
such vector fields. ``Infinitesimal'' changes of those parameters generate
what we intuitively think of as infinitesimal variations of the system.

Then, as discussed in the introduction, a phase space vector field
$\delta_t$ (equivalently an evolution of the system) is said to be
Hamiltonian if there exists a function $H_t$ such that
\be
\Omega( \delta_t, \delta) = \delta H_{t} \, ,
\ee
for all variations $\delta$. Conversely, given a Hamiltonian function
$H_t$, we can find out how it evolves the system by solving the above
equation for $\delta_t$. Thus, there is a mapping between Hamiltonian
evolutions and Hamiltonian functions.

As an example, for the evolution $\delta_t$ which gives rise to the
usual equations of motion:
\be
\frac{dq}{dt} = \frac{p}{m} \quad \mbox{ and } \quad
\frac{dp}{dt} = - kq, \label{pqdot}
\ee 
we can show that
\be
   \Omega(\delta_t, \delta) = \left(\frac{p}{m}\right) \delta p -
   (-kq) \delta q = \delta H_t,
\ee
where 
\begin{equation}
   H_t = \frac{p^2}{2m}+\frac{k q^2}{2} + C \, ,
\end{equation}
and $C$ is a free constant. Thus, the time evolution is generated by
$H_t$, which the reader will immediately identify as the classical
energy of the system (up to a constant). Note too that the equations
of motion confirm that $\delta_t H_{t} = 0$ (which we knew already by
the skew-symmetry of the symplectic structure).

\subsection{Time-dependent simple harmonic oscillator}
\label{s2.2}

Things get more complicated if $k$ is not a constant, but instead
varies in time. In this case, the energy of the oscillator will not be
constant, but instead may vary as a changing $k$ adds energy to or
removes it from the system (more properly, the external agency
setting $k$ can do net work on the system).  Thus, if we
have no knowledge of how $k$ is being fixed, we cannot do a standard
Hamiltonian analysis of the system -- such calculations require a
closed system which does not exchange energy with its surroundings.
However, we can use the extended phase space formalism to partially
compensate for our ignorance of the external system that sets $k$.
All we require is that $k(t)$ be a fixed function of $t$.

To allow for a time dependent spring constant $k$ and consequently a
time dependent energy, it is necessary to extend the phase space to
include $t$ and its conjugate momentum $p_{t}$.  The symplectic
structure is then given by
\begin{equation}
\label{sympart} 
   \Omega(\delta_1, \delta_2) =  
   (\delta_1 q)(\delta_2 p) - (\delta_2 q)(\delta_1 p)
   + (\delta_1 t)(\delta_2 p_t) - (\delta_2 t)(\delta_1 p_t).
\end{equation}

Furthermore, we would like to allow general evolutions, rather than
restricting to $(d/dt)$.  Thus, we shall study evolution generated by
\be
\Lambda = \lambda_o \frac{d}{dt} \, ,
\ee
where $\lambda_{o}$ is a free parameter which is required to be
strictly positive.  Our task is then to find a Hamiltonian %
    \footnote{We will follow the convention of \cite{arnold} and use
    $K$ to denote the Hamiltonian in extended phase space.}%
 $K_{\Lambda}$ which generates the following evolution:
\be
\label{variations}
\delta_\Lambda q = \lambda_{o} \frac{p}{m}, \quad
\delta_\Lambda p = - \lambda_{o} k q , \quad \mbox{and} \quad
\delta_\Lambda t = \lambda_o.
\ee
Note that the evolution of $p_{t}$ under $\delta_{\Lambda}$ is not
determined a priori.  This in turn leads to a freedom in the form of
the Hamiltonian $K_{\Lambda}$.  

We proceed by evaluating $\Omega(\delta_{\Lambda}, \delta)$.  Making
use of (\ref{variations}), as well as the fact that $k(t)$ is a fixed
function of $t$ so that 
\[ \delta k = \dot{k}\, \delta t \, \]
we obtain
\bea
   \Omega( \delta_\Lambda, \delta) &=& \lambda_o \,\delta \left( p_t +
     \frac{p^2}{2m} + \frac{k q^2}{2} \right) - \left(
     \frac{\lambda_{o} q^2 \dot{k}}{2} + \delta_{\Lambda}{p}_t \right)
     \delta t \nonumber \\   
   &=& \delta (\lambda_o K_{t}) - K_t \,\delta \lambda_o - \left(
     \frac{\lambda_{o} q^2 \dot{k}}{2} + \delta_{\Lambda}{p}_t \right)
     \delta t \, , \label{line2} 
\eea
where we have defined
\be
K_{t} := p_{t} + \frac{p^2}{2m} + \frac{k q^2}{2} \, .
\ee

Now, our evolution $\delta_\Lambda$ will be a Hamiltonian vector field
in the phase space if and only if $\Omega(\delta_\Lambda, \delta)$ is
an exact variation.  This will only be true if the $\delta
\lambda_{o}$ and $\delta t$ terms vanish. Therefore, we must be at a
point in phase space where
\begin{equation}\label{shoconstraint}
K_{t} = 0 \quad \Rightarrow \quad p_t = - \left(\frac{p^2}{2m} +
\frac{k q^2}{2}\right) \, . 
\ee
Furthermore, $p_{t}$ must satisfy the equation of motion:
\begin{equation}
\label{ptevol}
   \delta_{\Lambda} p_{t} = - \frac{\lambda_{o} q^2 \dot{k}}{2} \, ,
\end{equation}
which guarantees that the constraint (\ref{shoconstraint}) is
preserved under $\Lambda$-evolution. Then, on this constraint surface,
the evolution is Hamiltonian and generated by
\begin{equation}
K_\Lambda = \lambda_o K_{t}. \label{shoHam}
\end{equation}
The reader will immediately realize that this function vanishes
on-shell. However, this does not mean that the energy of the system
will vanish.  In the extended phase space, the energy is given by the
negative of the value of the momentum canonically conjugate to the
time.  Thus we obtain
\begin{equation}\label{energy}
   E_{t} := -p_{t} = \frac{p^2}{2m} + \frac{k q^2}{2} \, .
\end{equation}
This is immediately recognized as the usual energy associated to a
harmonic oscillator, although it will not necessarily be constant due
to the time dependence of $k(t)$.

The Hamiltonian $K_\Lambda$ given in (\ref{shoHam}) generates the
desired evolutions (\ref{variations}) of $t$, $p$, and $q$, but it is
by no means the unique Hamiltonian that does this. With no
$\delta_\Lambda p_t$ specified \textit{a priori}, the evolution of
$p_t$ can take any form that we like, and this freedom translates into
an ambiguity in both the Hamiltonian $K_\Lambda$ and the energy $E_t$.
We explore the range of possible Hamiltonians (and therefore energies)
by considering the functions that may be added to $K_\Lambda$ without
affecting the evolution equations (\ref{variations}).

To start one would consider functions of all possible variables and
parameters --- that is functions of the form $f(t,p_t, q, p,
\lambda_o)$. However, we immediately see that a dependence on $p_t$,
$q$, or $p$ will change the evolution equations
(\ref{variations}). Therefore only functions of the form
$f(t,\lambda_o)$ may be considered. For such functions, the constraint
equation $K_t = 0$ transforms to become
\be
\label{newconst}
p_t + E_t + \frac{\partial f}{\partial \lambda_o} = 0 \, .
\ee
Next, we demand that the constraint equations do not depend on the
Lagrange multiplier $\lambda_o$.  This is equivalent to requiring that
either the energy $E_t$ should not depend on $\lambda_o$ or equally
the equations of motion should not restrict the allowed values (or
evolution) of $\lambda_o$.  With this assumption, the freedom is
reduced to $f(\lambda_o,t) = \lambda_o g(t) + h(t)$, for any functions
$g(t)$ and $h(t)$.  Then, the derived equation of motion for $p_t$
(\ref{ptevol}) becomes
\be
\delta_\Lambda p_t = - \lambda_o \frac{\dot{k} q }{2} - 
\frac{\partial f}{\partial t} \, .
\ee
This will only be consistent with (\ref{newconst}) if $h(t)$ is in fact a 
constant $C$. Then any Hamiltonian of the form
\begin{equation}\label{generalshoham}
   K'_{\Lambda} = \lambda_{o} \left(p_{t} + \frac{p^2}{2m} + \frac{k
     q^2}{2} + g(t) \right) + C \, ,
\end{equation}
will be consistent with our requirements, and so the energy will only be
defined up to a free function:
\begin{equation}\label{genenergy}
   E'_{t} = \frac{p^2}{2m} + \frac{k q^2}{2} + g(t) \, .
\end{equation}

We finish this analysis of the freedom using a physical argument.
Mathematically, any function $g(t)$ will satisfy our requirements.
Physically however, it is reasonable to demand that this function
should be in some way connected to the system. With that requirement
we are reduced to considering functions of $k(t)$.

\section{Gravity}
\label{s3}

We now turn to a canonical Hamiltonian formulation of general
relativity over a quasi-local region of space-time. Although the
technical details will, of course, be much more complicated, many of
the basic conceptual issues relating to extended phase space have
already been dealt with in the previous example. Thus, for gravity we
will also extend the usual phase space to include a time variable ---
which will be defined only on the boundary --- and it will be joined by
the spatial coordinates of the boundary. Further, just as $k$ in the
above was only fixed up to changes in the time parameter, for
quasi-local gravity our boundary conditions will only be fixed up to
intrinsic diffeomorphisms of that boundary. We will also find that a
range of Hamiltonians will generate the standard evolutions of a
space-time and that this freedom may be traced to the freedom to choose
the evolution of the conjugate momenta of the boundary coordinates.
Each of these Hamiltonians will be valid on its own constraint
surface, and again each of these constraint surfaces will correspond
to a different energy function for the system.

\subsection{Manifold without boundary}
\label{s3.1}

Let us begin by briefly reviewing the Hamiltonian formulation of
general relativity for a manifold with no boundaries.  This will also
serve to fix our notation and conventions.  We will consider
time-dependent fields living on a space-like three-manifold $\Sigma$.
The intrinsic geometry of $\Sigma$ is fully specified by a space-like
three-metric $h_{ab}$; the derivative operator compatible with the
metric will be denoted by $D_{a}$.  In the standard Hamiltonian
formulation, the 3-metric $h_{ab}$ serves as the configuration
variable, while its conjugate momentum is the tensor density $P^{ab}$.
Thus, the phase space consists of pairs of fields $(h_{ab}, P^{ab})$
with the symplectic structure
\begin{equation}\label{noboundss}
   \Omega(\delta_{1}, \delta_{2}) = \int_{\Sigma} d^{3}x \, \left\{
   (\delta_{1} h_{ab}) (\delta_{2} P^{ab}) - (\delta_{2} h_{ab})
   (\delta_{1} P^{ab}) \right\} \, .
\end{equation}

To specify a time evolution on $\Sigma$, we introduce a lapse function
$N$ and a shift vector field $V^a \in T \Sigma$. In the usual way the
lapse and the shift will prescribe how time ``flows'' on
$\Sigma$. Only after these fields are given can we define a time
derivative $\frac{d}{dt}$ over $\Sigma$, as only then will we know how
to associate points at a time $t$ with points at time $t + \delta t$,
and also know how much proper time has passed during that interval.
Given a flow of time (or equivalently a lapse and shift), we introduce
a Hamiltonian
\begin{equation}\label{noboundham}
   H_{t} = \int_{\Sigma} d^{3}x \left\{N\mathcal{H} + V^{a}
   \mathcal{H}_{a} \right\}
\end{equation}
which generates time evolution.  Specifically, we obtain
\bea
\frac{d}{dt} {h}_{ab} &=& [h_{ab}]_{(N,V)} \label{hdot} \mbox{ and} \\ 
\frac{d}{dt} {P}^{ab} &=& [P^{ab}]_{(N,V)} \label{Ghh} \, ,
\eea
where the exact forms of $[h_{ab}]_{(N,V)}$ and $[P^{ab}]_{(N,V)}$ are given
(along with expressions for the Hamiltonian constraint $\mathcal{H}$
and diffeomorphism constraint $\mathcal{H}_a$) in \ref{messy}.
Furthermore, the initial data for $h_{ab}$ and $P^{ab}$ must satisfy
the constraints 
\be\label{constraints} 
\mathcal{H} = 0  \; \; \mbox{ and} \; \; \mathcal{H}_a = 0 \, .
\ee
which are then automatically preserved in time.

\subsection{Boundary Conditions}
\label{s3.2}

We would like to extend the Hamiltonian formulation of general
relativity to manifolds with boundary.  Therefore, we will now
consider $\Sigma$ to be a space-like 3-manifold with a closed
2-boundary $\mathcal{B}$.  In this section, we will describe the
boundary conditions enforced on $\mathcal{B}$.  The essential idea is
to keep the boundary metric, lapse and shift fixed, up to
diffeomorphisms of the boundary.  To make this precise, we proceed as
follows.

Construct a three-manifold $B \cong \mathcal{B} \times \Rbar$ which is
foliated by two-manifolds $\mathcal{B}_{\tid{t}} \cong \mathcal{B}$
where $\tid{t}$ is the foliation parameter and ``$\cong$'' indicates
that the manifolds are diffeomorphic.  The parameter $\tid{t}$
provides a notion of time on $B$.  Specifically, the
$\mathcal{B}_{\tid{t}}$ are taken as ``instants'' of time, and
$\tid{t}_2$ is said to occur after $\tid{t}_1$, if $\tid{t}_2 >
\tid{t}_1$.  We then introduce a time-like three-metric
$\tid{\gamma}_{ab}$ on $B$ which pulls back to a space-like two-metric
$\tid{\sigma}_{ab}$ on each of the elements of the foliation.  Using
the three-metric, we can also obtain the future-directed unit normal
vector field to the $\mathcal{B}_{\tid{t}}$, which we denote
$\tid{u}^{a}$.  Finally, we introduce a time evolution vector field
$\tid{\part}^a \in TB$ which satisfies
\be
\label{Tdt}
\Lie_{\tid{T}} \tid{t} = 1 \, .
\ee
We note that $\tid{T}^a$ may be decomposed into its parts
perpendicular and parallel to the $\mathcal{B}_{\tid{t}}$ as
\be
\tid{T}^a = \tid{N} \tid{u}^a + \tid{V}^a \, ,
\ee
where $\tid{N}$ is the lapse function and $\tid{V}^a \in T
\mathcal{B}_{\tid{t}}$ is the shift vector on $B$. Then, the future
pointing unit normal to $\mathcal{B}_{\tid{t}}$ is
\be
\tid{u}_a = - \tid{N} d\tid{t}_a \, .
\ee
We can also decompose the metric $\tid{\gamma}_{ab}$ as
\begin{equation}
\tid{\gamma}_{ab} = - \tid{N}^2 d\tid{t}_{a} d\tid{t}_{b} 
+ \tid{\sigma}_{ab} \, ,
\end{equation}
while
\be
\tid{\gamma}^{ab} = - \frac{1}{\tid{N}^2} \tid{T}^a \tid{T}^b 
+ \frac{1}{\tid{N}} \tid{T}^{(a}\tid{V}^{b)} 
+ \tid{\sigma}^{ab}.
\ee

With this framework in place, we are ready to introduce our
boundary condition on $\mathcal{B}$. 

\medskip \noindent \textbf{The Boundary Conditions}

\begin{enumerate}

\item  Construct the 3-manifold $B$ and equip it with a foliation,
  fixed time-like boundary metric $\tid{\gamma}_{ab}$ and time
  evolution vector field $\tid{T}^{a}$ as described above.

\item  Introduce a smooth diffeomorphism
\be
\omega: \mathcal{B} \times \Rbar \rightarrow B,
\ee
so that if $t$ is the time parameter for $\Sigma$ and $\mathcal{B}$,
and $\tid{t}(t)$ is some monotonically increasing function from $\Rbar
\rightarrow \Rbar$, then
\be
\omega( \cdot, t) : \mathcal{B} \rightarrow \mathcal{B}_{\tid{t}(t)},
\ee
is a diffeomorphism for all $t \in \Rbar$ (see figure \ref{omega}).
That is, instants of time on $\mathcal{B}$ are mapped to our 
pre-defined ``instants'' of time in $B$.

\item The two-metric $\sigma_{ab}$, lapse $N$, and shift $V^a$ on
$\mathcal{B}$ at a time $t$ are equal to the corresponding
$\tid{\sigma}_{ab}$, $\tid{N}$, and $\tid{V}^a$ pulled-back to
$\mathcal{B}$ using $\omega (\cdot, t)$ (or pushed-forward using
$\omega^{-1} (\cdot, \tid{t} )$ in the case of $\tid{V}^a$).

\item The extrinsic curvature induced on $B$ by $h_{ab}$ and $P^{ab}$
must satisfy the time-like diffeomorphism constraint.

\end{enumerate}

\begin{figure}
\center{\resizebox{!}{2in}{\includegraphics{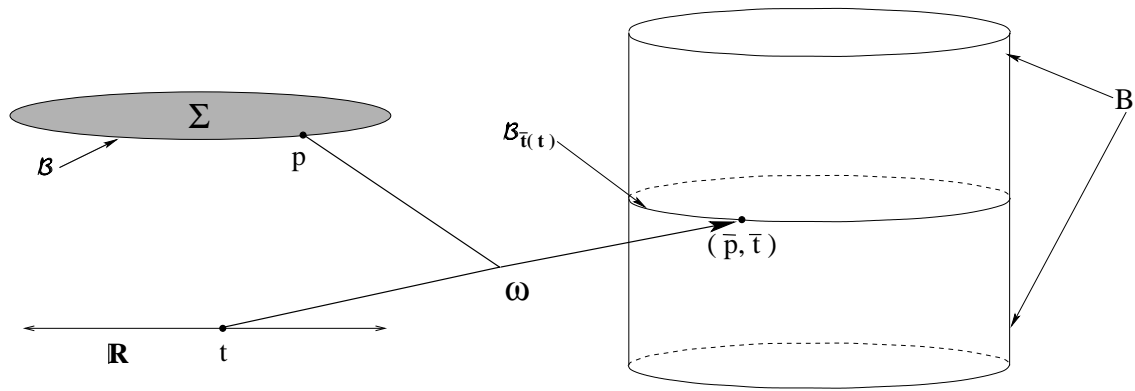}}}
\caption{The map $\omega$ (with one dimension of $\mathcal{B}$
  suppressed). } 
\label{omega}
\end{figure}

In condition (iii), we have required $V^a = \tid{V}^{a}$ which ensures
$V^a \in T \mathcal{B}$ on $\mathcal{B}$. However, in the initial
set-up, we only required that $V^a \in T \Sigma$. This extra
restriction is equivalent to the orthogonality assumption of
Brown-York (which is discussed in more detail in section \ref{s4}) and
is not essential but will somewhat simplify the already involved
discussion that will follow. The more general non-orthogonal case,
which only requires that $N^2 - V^a V_a = \tid{N}^2 - \tid{V}^a
\tid{V}_a$ and is essential if we wish to allow $B$ to be null or
space-like, has been studied and will be discussed in a future
paper. For now, however, we will work with this orthogonality
assumption and so find that
\be
\frac{d}{dt} = \Lie_{\tid{T}} \, ,
\ee
when these operators act on fields that are defined on $B$ and then
mapped back to $\mathcal{B}$.   Finally, (iv) says that even when we
later consider general variations of our fields, they must continue to
satisfy the diffeomorphism constraint at $\mathcal{B}$.

In order to describe the phase space of quasi-local general relativity,
we need a concrete realization of the diffeomorphism $\omega$. To
obtain this, we provide a coordinatization of the three-manifold $B$
in terms of the foliation parameter $\tid{t}$ and two ``angular''
coordinates $\tid{x}^A$ on $\mathcal{B}_t$
\footnote{In many cases, multiple coordinate
systems will be required to cover $B$ without coordinate
singularities and corresponding singularities in the coordinate vector
fields. The calculations of this paper extend directly to cover these
situations, but in the interests of clarity, we'll proceed as if one
set of angular coordinates was sufficient.},
which are chosen to be ``time-independent'' --- that is 
\be
\Lie_{\tid{\part}} \tid{x}^{A} = 0.
\label{dxdt}
\ee
(Here, and in the remainder of the paper, we will use capital letters
to signify the coordinates and components of tensors in this chosen
coordinate system, while lower case letters are the usual abstract
index notation.) Once the coordinate system on $B$ is given, $\omega$
may be described by how it pulls back the coordinate system to
$\mathcal{B} \times \Rbar$.  Giving $(\tid{x}^{A}, \tid{t})$ at each
point in $\mathcal{B} \times \Rbar$ uniquely determines the map
$\omega$.  Thus, we can (and do) specify $\omega$ by assigning an
$\tid{x}^{A}$ and $\tid{t}$ to each time and place in $\mathcal{B}$.

Notice that while (ii) specifies that a diffeomorphism exists, the
exact form of $\omega$ is not fixed.  However, any permissible
$\omega$ can be obtained from any other
\footnote{Up to potential global topological obstructions which we
shall ignore with impunity since in our calculations we will only
consider ``infinitesimal'' variations of $\omega$.}
simply by composing $\omega$ with a diffeomorphism
\be
   \phi: B \rightarrow B \quad \mbox{ mapping } \quad 
         (\tid{p},\tid{t}) \mapsto (\tid{p}',\tid{t}') 
= \phi(\tid{p},\tid{t}) \, , \nonumber 
\ee
which preserves the foliation of $B$.  Thus, the freedom in defining
$\omega$ is the freedom to consider a preferred $\omega_{o}$ composed
with all $\phi$, where 
\be
\phi \circ \omega_{o}: \mathcal{B} \times \Rbar \rightarrow B 
\quad \mbox{maps} \quad 
(p,t) \mapsto \phi \left(\omega_{o}(p,t) \right) \nonumber \, .
\ee
Since $\omega_{o}$ mapped all fields ($\tid{\sigma}_{ab}$, $\tid{N}$,
$\tid{V}^a$, and the coordinates $\tid{t}$ and $\tid{x}^A$) back to
$\mathcal{B}$, then $\phi \circ \omega_{o}$ will too. That is, from a
computational point of view, our diffeomorphisms act so that observers
fixed to points either on $B$ or $\mathcal{B}$ will see all of these
fields perturbed.

For an infinitesimal diffeomorphism $\phi$, it is straightforward to
calculate the changes to the various fields at the boundary.  We begin
by recalling that infinitesimal diffeomorphisms are generated by
non-singular vector fields and so to each $\phi$ we may associate a
vector field $\delta X = (\delta \tid{t}, \delta \tid{x}^a)$.  The
requirement that $\phi$ preserve the foliation means that $\delta
\tid{t}$ should be a constant on each leaf of the foliation, although
it can vary from one leaf to the next. Meanwhile, $\delta \tid{x}^a$
can be any non-singular vector field in $TB$ that is everywhere
parallel to the foliation surfaces. Under the infinitesimal action of
this diffeomorphism the fields on the boundary $B$ will change
according to
\begin{eqnarray}
   \tid{\gamma}_{ab} &\mapsto& \tid{\gamma}_{ab} + \Lie_{\delta X}
   \tid{\gamma}_{ab} 
   \nn \\  
   d\tid{t} &\mapsto& d\tid{t} + \Lie_{\delta X} d\tid{t}
   \nn \\ 
   \tid{\part} &\mapsto& \tid{\part} + \Lie_{\delta X}
   \tid{\part} \nn \, \\ 
   \tid{t} &\mapsto& \tid{t} + \delta \tid{t} \quad
   \mbox{and} \quad \nn \\ 
   \tid{x}^A &\mapsto& \tid{x}^A + \delta \tid{x}^A. \nn
\end{eqnarray}
Here, $\delta \tid{x}^{a}$ is simply the vector field $\delta
\tid{x}$, expressed using the abstract index notation, while the $\delta
\tid{x}^{A}$ are the \textit{components} of the vector field $\delta
\tid{x}$.  It will be important to keep this distinction clear.  The
two are related by
\begin{equation}\label{components}
   \delta \tid{x}^A = (\delta \tid{x})^{a} d_a \tid{x}^{A} \, ,
\end{equation}
where $d_a$ is the intrinsic derivative over $\mathcal{B}$.

The corresponding change to the fields in $\mathcal{B} \times \Rbar$
is given by
\begin{eqnarray}
   N &\mapsto& N + (\delta t) \frac{d}{dt} N + \Lie_{\delta x} N
      \label{dN} \\  
   V^a &\mapsto& V^{a} +(\delta t) \frac{d}{dt} V^a + \Lie_{\delta x}
      V^a \label{dV} \\  
   \sigma_{ab} &\mapsto& \sigma_{ab} + (\delta t) \frac{d}{dt}
      \sigma_{ab} + \Lie_{\delta x} \sigma_{ab} \label{dsig} 
      \, , \\  
   t &\mapsto& t + \delta t \quad \mbox{ and} \label{dt} \quad \\ 
   x^A & \mapsto & x^A + \delta x^A, \label{dx}
\end{eqnarray}
after $\omega$ maps everything back to $\mathcal{B}$. Note that we
have dropped the over-bars in order to reduce notational
clutter. In the future we will blur the distinction between
$\mathcal{B} \times \Rbar$ and $B$. Whenever there is an ambiguity,
$\omega$ is understood to be acting to make the identification and map
quantities back and forth.

To summarize, our boundary conditions imply that once a map $\omega$
from $\mathcal{B} \times \Rbar$ to $B$ is specified, the boundary
metric, lapse and shift are known.  We have simplified matters by
introducing a coordinate system on $B$ which allows us to easily
characterize the map $\omega$, but the results obtained in the
following subsections will not be sensitive to this coordinate system,
and it is likely that the calculations could be done without
introducing coordinates at all.  What will be important is that the
\textit{only} allowed variations of the boundary will be infinitesimal
diffeomorphisms.  They will generate changes in the boundary metric,
lapse and shift as given by (\ref{dN}-\ref{dsig}).  Additionally, they
will change the map $\omega$, or equivalently the coordinates
associated to the points of $\mathcal{B}$, according to
(\ref{dt},\ref{dx}).

Comparing with earlier work, Regge and Teitelboim \cite{rt} showed that their
Hamiltonian was covariant with respect to the Poincar\'{e} group at infinity
acting on the boundary. Here we have set things up so that we may study the
effect of the diffeomorphism group that maps the boundary into itself 
(while preserving the foliation) on the Hamiltonian. Thus, apart from 
the difference between boundaries at finite difference and infinity, 
we are also studying different group actions on those boundaries.

\subsection{Phase Space and Hamiltonian Evolution}
\label{s3.3}

In the previous subsections, we considered the fields in the bulk as
well as the boundary conditions imposed at $\mathcal{B}$. In this
subsection, we will turn our attention to the phase space and
associated symplectic structure, as well as the Hamiltonians
associated with evolution equations.

With the boundary conditions that we have imposed, the configuration
variables of a point in the phase space will be given by the
three-metric $h_{ab}$, a value of the time parameter $t$, and a set of
angular coordinates $x^A$ on $\mathcal{B}$.  The conjugate momenta to
these configuration variables will be $P^{ab}$, $P_t$, and $P_A$, and
so the coordinates of a point in phase space will be given by the six
fields $(h_{ab}, P^{ab}, t, P_{t}, x^{A}, P_{A})$. Note that our
boundary conditions say that once we specify $t$ and the $x^A$, we
will know $\sigma_{ab}$ over $\mathcal{B}$, thus $h_{ab}$ cannot be
chosen completely independently of those variables.  The symplectic
structure %
\footnote{This is really the pre-symplectic structure since the
existence of the Hamiltonian and diffeomorphism constraints means that
we have not properly isolated the true degrees of freedom. The proper
symplectic structure would restrict itself to these real degrees of
freedom. That said, we will follow the standard practice and work with
the pre-symplectic structure.} 
on the phase space then takes the form:
\bea \label{sympl}
   \Omega(\delta_1,\delta_2) &=& \int_\Sigma d^3 x \left\{ (\delta_1
     h_{ab})(\delta_2 P^{ab}) - (\delta_2 h_{ab})(\delta_1 P^{ab})
     \right\} \\ 
   &&\quad + \int_{\mathcal{B}} d^2 x \left\{ (\delta_1 x^A)(\delta_2
     P_A) - (\delta_2 x^A)(\delta_1 P_A) \right\} \nn \\ 
   &&\quad + \left\{(\delta_1 t)(\delta_2 P_t) - (\delta_2 t)(\delta_1
     P_t) \right\} \, . \nonumber
\eea

The variations which appear in the bulk are entirely free, and are not
restricted to being on-shell.  However, they are partially restricted
at the boundary. Specifically, variations must preserve the boundary
conditions, and so must be generated by some $(\delta t, \delta x^A)$.
Then the variations of the lapse $N$, shift $V^a$, two-metric
$\sigma_{ab}$, the time parameter $t$, and the coordinates $x^{A}$ are
given by equations (\ref{dN}-\ref{dx}).

A typical on-shell variation $\delta_\Lambda$ will be generated by the
action of
\be 
   \Lambda = \lambda_{o} \frac{d}{dt} + \Lie_\lambda , 
\label{Lambda}
\ee 
where $\lambda_{o}$ must be constant on the boundary (so as to
preserve the foliation of the boundary) and $\lambda \in T \Sigma_t$
in the bulk while it is tangent to $\mathcal{B}_t$ on the boundary.
Then we know how our phase space variables must evolve with
$\Lambda$. Namely,
\bea
\label{hpevol}
\delta_\Lambda h_{ab} &:=& 
\lambda_o \frac{d}{dt} h_{ab} + \Lie_\lambda h_{ab}, \\
\label{Pevol}
\delta_\Lambda P^{ab} &:=& 
\lambda_o \frac{d}{dt} P^{ab} + \Lie_\lambda P^{ab}, \\
\label{dtdL} \delta_\Lambda t &:=& \lambda_o, \quad \mbox{ and} \\
\label{dxdL} \delta_\Lambda x^A &:=& \lambda^a d_{a} x^{A} \, , 
\eea
where the time derivative of $h_{ab}$ and $P^{ab}$ is given in
(\ref{hdot}) and (\ref{Ghh}) respectively. The evolution of the
remaining two fields, $P_{t}$ and $P_{A}$ is currently undetermined.

We would now like to determine whether there is a Hamiltonian
$K_\Lambda$ which generates the on-shell evolution
$\delta_{\Lambda}$ given in equations (\ref{Lambda}-\ref{dxdL}).  Thus
we will try to manipulate the symplectic structure into the form
\[ \Omega(\delta_\Lambda,\delta) = \delta K_{\Lambda} +
\mathrm{constraints}. \]
In doing this we will be aided by the fact that $\delta_{\Lambda}
P_{t}$ and $\delta_{\Lambda} P_{A}$ are as yet unspecified.

The first important step in this process is to evaluate the bulk term
in the symplectic structure.  Here, we make use of the result of
\ref{appb}:
\bea\label{dH} 
&&\int_\Sigma d^{3}x \left\{ (\delta_\Lambda h_{ab})(\delta P^{ab}) -
   (\delta_\Lambda P_{ab})(\delta h_{ab}) \right\} = \\
&&\quad \delta \left(\int_{\Sigma}d^{3}x \left\{ \lambda_{o}\, N \, \cH
   + (\lambda^a + \lambda_{o} V^{a})\cH_{a} \right\} 
   + \int_{\mathcal{B}} d^{2}x \sqrt{\sigma}
   \left[\lambda_{o}(N\varepsilon - V^{a} j_{a}) - \lambda^{a} 
   j_{a}\right] \right) \nonumber \\
&& \quad - \int_{\Sigma} d^{3}x \left\{\delta(\lambda_{o}\, N) \cH +
   \delta(\lambda^{a} + \lambda_{o} V^{a})\cH_{a} \right\} \nonumber \\  
&& \quad - \int_\mathcal{B} d^{2}x \sqrt{\sigma} \left[ \varepsilon
   \delta (\lambda_{o}N) - j_{a} \delta(\lambda_{o} V^a + \lambda^a)  - 
   \frac{\lambda_{o} N}{2} s^{ab} \delta \sigma_{ab} \right]. \nn
\eea
As one would expect from its general appearance, the derivation of
this result is fairly involved. It is discussed in greater detail in
the appendix.  Here however, we simply note that
\bea\label{ejs}
\varepsilon & := & k/(8 \pi G), \nonumber \\
j_a & := & - 2 \sigma_{ac} P^{cd} n_d /\sqrt{h}, \,\,
\mbox{and}\nonumber \\
s^{ab} & := & (1/8 \pi G) \left( k^{ab} - (k - n^c a_c) \sigma^{ab}
\right), 
\eea
where $k = - \sigma^{ab} D_a n_b$ is the extrinsic curvature of
$\mathcal{B} \in \Sigma$ (note the sign convention and recall that
$D_a$ is the covariant derivative in $\Sigma$ that is compatible with
$h_{ab}$), $a_c = \frac{1}{N} D_c N$ and $G$ is Newton's constant. The
meanings of some of these quantities are easier to see if one thinks
of them as being defined in four-dimensional space-time --- this
perspective is discussed in section \ref{s4} following equation
(\ref{gravHam}).

In the last line of (\ref{dH}), there are terms involving the
variation of the boundary 2-metric, lapse and shift.  However, due to
our boundary conditions, we know that $\delta$ must act on the lapse,
shift, and two-metric like an infinitesimal diffeomorphism in $B$
generated by the vector field 
\[ \delta X^a = \delta t \,\part^a + (\delta x)^{a} \, .\]
Then, with the help of \ref{calc}, we see that because the equations
of motion hold at the boundary,
\bea\label{eq2} 
   && \int_{\mathcal{B}_t} d^{2}x \, \ssg \left( \varepsilon
      \Lie_{\delta X} N -
      j_a \Lie_{\delta X} V^a - \frac{N}{2} s^{ab} \Lie_{\delta X}
      \sigma_{ab} \right) =  \nonumber \\ 
   &&\qquad \int_{\mathcal{B}_t} d^{2}x \, (\delta t) \, \frac{d}{dt}
      (\ssg [N \varepsilon - V^a j_a]) 
   - \int_{\mathcal{B}_t} d^{2}x \, (\delta x)^a \, \frac{d}{dt} (\ssg
   j_a) \, ,
\eea
where the four-dimensional, in-boundary Lie derivatives $\Lie_T$ from
\ref{calc} are replaced with $\frac{d}{dt}$ from our three-dimensional
perspective.  We can now use (\ref{dH}) and (\ref{eq2}) to rewrite the
bulk part of the symplectic structure in (\ref{sympl}) to obtain
\bea
\label{symplectic}
\Omega(\delta_\Lambda, \delta) & = & 
\delta \left( \int_{\Sigma} d^{3}x \left[ (\lambda_{o}N)\cH +
   (\lambda_{o}V^{a} + \lambda^{a}) \cH_{a} \right] \right) \nonumber \\
&&+ \delta \left( \int_{\mathcal{B}} d^{2}x \, \sqrt{\sigma}
   \left[\lambda_{o}(N\varepsilon - V^{a}\, j_{a}) - \lambda^{a}\, j_{a}
   \right] + \lambda_{o}P_{t} + \int_{\mathcal{B}} d^{2}x \,
   \lambda^{A} P_{A} \right) \nonumber \\
&&- \int_{\Sigma} d^{3}x \, \left[ \delta(\lambda_{o}N) \cH +
   \delta(\lambda_{o} V^{a} + \lambda^{a})\cH_{a} \right] \nonumber \\
&&- (\delta \lambda_{o}) \left[ P_{t} + \int_{\mathcal{B}} d^{2}x
   \sqrt{\sigma} \left(N \varepsilon - V^{a} j_{a} \right) \right]
   \nonumber  \\
&&- (\delta t)  \left[ \delta_{\Lambda} P_{t} + \int_{\mathcal{B}}
   d^{2}x \, \lambda_{o} \frac{d}{dt} \left(\sqrt{\sigma}(N \varepsilon -
   V^{a} j_{a}) \right) \right] \nonumber \\
&&- \int_{\mathcal{B}} d^{2}x \left[\,(\delta \lambda^{A}) P_{A} -
   (\delta \lambda^a) \sqrt{\sigma} j_{a}\right] \nonumber \\
&&- \int_{\mathcal{B}} d^{2}x \, (\delta x)^A \left[ \delta_{\Lambda}
P_{A} - \lambda_{o} \, \frac{d}{dt} \left(\sqrt{\sigma} j_{A} \right) 
\right] \, .
\eea

This is nearly the desired form.  We see that the first two lines are
an exact variation, as desired.  The third line will vanish provided
the usual bulk constraints of general relativity are satisfied.  The
fourth line is then a new constraint which appears at the boundary and
relates $P_t$ to other fields on the boundary.  The fifth line gives
the evolution of $P_{t}$ and guarantees that the constraint will
continue to hold (these two correspond to the constraint and evolution
equation found for $p_t$ in the case of the harmonic oscillator).  We
would like to write the final two lines in a similar form --- a
constraint for $P_{A}$ and an evolution equation which ensures that
this constraint is preserved.  To this end, we must evaluate $\delta
(\lambda^{A})$.  Making use of (\ref{components}), it follows that
\begin{equation}
   \delta (\lambda^A)= (\delta \lambda)^{a}(d_{a}x^{A}) - \lambda^{a}
   d_{a} (\delta x^A) \, , 
\end{equation}
where as usual the capital Latin index indicates a component of a
field in the coordinate system imposed on $\mathcal{B}$, while the lower
case index is an abstract tensor index.  Therefore, we obtain:
\be
\label{deltalamrel}
\delta (\lambda^A) P_A - (\delta \lambda)^a \ssg j_a 
= (P_A - \ssg j_A) \delta (\lambda^A) - \ssg j_A \lambda^a d_a (\delta
x^A) \nn \, , 
\ee
where $j_A = j_a \left[ \frac{\partial}{\partial x^A} \right]^a$.
Substituting this expression into (\ref{symplectic}), doing some
integration by parts, and using Stokes theorem, we can rewrite 
the last two lines as
\begin{equation}\label{last2lines}
- \int_{\mathcal{B}} d^{2}x (\delta \lambda^A) \left[P_{A}
   -  \sqrt{\sigma} j_{A}\right] 
   - \int_{\mathcal{B}} d^{2}x \, (\delta x)^A \left[ \delta_{\Lambda}
   P_{A} - \delta_\Lambda (\sqrt{\sigma} j_{A}) \right] \, ,
\end{equation}
where as usual $\delta_\Lambda = \lambda_{o} \frac{d}{dt} +
\Lie_\lambda$. 

In order to simplify matters, let us introduce some notation:
\bea
   H_\Lambda & = & \int_\Sigma d^3 x \left\{ \lambda_{o} (N \cH + V^a
     \cH_a) + \lambda^a \cH_a \right\}  \\ 
   K_{t} &=& P_{t} + \int_{\mathcal{B}} d^2 x \ssg [ N \varepsilon -
     V^a j_a ] \,\, \mbox{and} \label{kt} \\
   L_A &=& P_A - \ssg j_A \, .
\label{Jx} \eea

Making use of the new notation, as well as (\ref{last2lines}), we can
rewrite the symplectic structure (\ref{symplectic}) as
\begin{eqnarray}\label{finalform} 
\Omega(\delta_\Lambda, \delta) & = & 
\delta \left(H_{\Lambda} + \lambda_{o} K_{t} + \int_{\mathcal{B}}
   d^{2}x \, \lambda^{A} L_{A} \right) \nonumber \\ 
&& - \int_{\Sigma} d^{3}x \, \left[ \delta(\lambda_{o}N) \cH +
   \delta(\lambda_{o} V^{a} + \lambda^{a})\cH_{a} \right] \nonumber \\
&& - (\delta \lambda_{o}) K_{t} - \int_{\mathcal{B}} d^{2}x (\delta
   \lambda^{A}) L_{A} \nonumber \\
&& - (\delta t) (\delta_{\Lambda} K_{t}) - \int_{\mathcal{B}} d^{2}x
   (\delta x)^A (\delta_{\Lambda} L_{A}) \, .
\end{eqnarray}
In order for a Hamiltonian generating evolution along $\Lambda$ to
exist, the right hand side of (\ref{finalform}) must be an exact
variation.  Thus, all terms after the first --- which is already an
exact variation --- must vanish.  This can be accomplished if the last
three lines vanish.Therefore, the evolution along $\Lambda$ will be
generated by a Hamiltonian provided that we are ``on-shell,'' by which
we mean:
\begin{eqnarray}\label{constrain} 
   \cH &=& 0 \quad \mbox{and} \quad \cH_{a} = 0 \, ; \nonumber \\
   K_{t} &=& 0 \quad \mbox{and} \quad L_{A} = 0 \, .
\end{eqnarray}
The first two expressions are the usual constraints of general
relativity, while the last two restrict our evolution to a constraint
surface in the extended phase space.  They can equivalently be thought
of as fixing the values of $P_{t}$ and $P_{A}$ to be
\begin{equation}\label{boundconst}
    P_{t} = - \int_{\mathcal{B}} d^{2}x \, \ssg [ N \varepsilon - V^a
    j_a ] \quad \mbox{and} \quad P_A = \ssg j_A \, .
\end{equation}
Finally, the last two terms in (\ref{finalform}) define the action of
$\delta_\Lambda$ on $P_{t}$ and $P_{A}$.  Essentially, it fixes the
evolution so that it evolves points on the constraint surface defined
by equations (\ref{boundconst}) into other points on that surface.

Therefore, we have shown that ``on-shell,'' i.e. when
(\ref{constrain}) is satisfied, the evolution along the vector field
$\Lambda$ is generated by the Hamiltonian $K_{\Lambda}$ which is 
given as: 
\begin{equation}\label{gravham}
   K_{\Lambda} = \int_{\Sigma}d^{3}x \left\{ (\lambda_{o} N)  \cH +
   (\lambda^a + \lambda_{o} V^{a})\cH_{a} \right\} + \lambda_{o}
   K_{t} + \int_{\mathcal{B}} d^{2}x \, \lambda^A L_A \, . 
\end{equation}
Although we have introduced a coordinate system on the boundary in
order to characterize the diffeomorphism $\omega$, the final form of
the Hamiltonian is independent of this choice.  Specifically, the term
$\lambda^{A} L_{A}$ will be the same when evaluated in any set of
coordinates.  As in the previous example, this Hamiltonian is not
determined uniquely.  However, we shall postpone discussion of the
ambiguities to the next subsection.

For the spherically symmetric case, a similar analysis to this one may
be found in \cite{kuchar}. In that case, the symmetry means that there
is no need to include ``angular" coordinates as configuration
variables.

\subsection{Energy and Angular Momentum}
\label{s3.4}

The on-shell value of the Hamiltonian (\ref{gravham}) generating
evolution along $\Lambda$ will be zero.  To find the energy and
angular momentum we must again consider the value of the
conjugate momenta rather than the Hamiltonian.  Since the equations of
motion guarantee that $K_{t}$ vanishes on-shell, we can use (\ref{kt})
to find the value of $P_{t}$.  The energy associated to time
translation is then simply the negative of this:
\begin{equation}\label{et}
   E_{d/dt} := - P_{t} = \int_{\mathcal{B}} d^{2}x
   \, \sqrt{\sigma}\left(N \varepsilon + V^a j_a \right)\, , 
\end{equation}
Similarly, the equations of motion show that $L_{A}$ vanishes on
shell, so that 
\begin{equation}\label{jt}
   P_{A} = \sqrt{\sigma} j_{A}, 
\end{equation}
Combining these two results, we arrive at the expression
\begin{equation}\label{elambda}
   E_{\Lambda} = \int_{\mathcal{B}} d^{2}x \, \left\{ \lambda_o  
   (N \varepsilon + V^a j_a)  + \lambda^a j_a \right\} \, , 
\end{equation}
for the ``charge'' associated with evolution according to
$\Lambda$. Specializing to the case of pure time evolution, we obtain
(\ref{et}) which is the energy associated with the time translation
$\frac{d}{dt}$.  Similarly, given a vector field $\phi$ tangent to the
3-surface $\Sigma$ and tangent to $\mathcal{B}$ at the boundary, the
angular momentum is given by
\begin{equation}\label{angmom}
   J_{\phi} = \int_{\mathcal{B}} d^{2}x \sqrt{\sigma} \phi^{a} j_{a}
   \, .
\end{equation}
As in the case of the Hamiltonian, our expressions for the energy and
angular momentum associated to the boundary were obtained by making
use of a specific coordinatization of the boundary.  However, the
final results are independent of the choice of coordinates, as one
might have hoped.

Finally, we turn to the ambiguity in the Hamiltonian $K_{\Lambda}$ 
and the corresponding freedom in the definition of the energy and angular 
momentum associated to the boundary. We follow the same procedure that
we used in section \ref{s2.2} to explore the allowed freedom in the 
Hamiltonian for the simple harmonic oscillator. Namely, we start by
considering adding any functional of the form
\be
F = \int_{\mathcal{B}} d^2 x f(h_{ab},P^{ab},t,P_t,x^A,P_A,\lambda_o, 
\lambda^A) \, ,
\ee
to $K_\Lambda$.
However, on requiring that:
\begin{enumerate}
\item the new Hamiltonian still generates the same evolution equations
(\ref{hpevol}-\ref{dxdL}),
 
\item the new versions of the constraints $K_t = 0$ and $L_A = 0$
remain independent of the Lagrange multipliers $\lambda_o$ and
$\lambda^A$ (equivalently either energy and angular momentum are
independent of these parameters or $\lambda_o$ and $\lambda^A$ may be
freely chosen without reference to the equations of motion or
constraints), and

\item the new versions of these constraints are preserved by the new
evolution equations for $P_t$ and $P_A$,

\end{enumerate}

the freedom is greatly reduced. It turns out that subject to these
conditions, the only valid Hamiltonian functionals will take the form,
\be
K'_\Lambda = K_\Lambda + \lambda_o F(t) + C \, ,
\ee
where $F(t)$ is a free function and $C$ is a free constant.

Furthermore, if we again make the argument that the only
physically relevant functions $F(t)$ are those which are associated
with the boundary data, then these free functions must take the form
\be\label{boundterm}
   F(t) = \int_\mathcal{B} d^2 x \ssg f(\sigma_{ab}, N, V^a), 
\ee
where $f$ is a free function of the given fields.  Therefore, we
find that the energy associated to an evolution along $\Lambda$ is
\begin{equation}\label{elambdafinal}
   E'_{\Lambda} = \int_{\mathcal{B}} d^{2}x \, \left[ \lambda_o  
   (N \varepsilon + V^a j_a)  + \lambda^a j_a \right] +
   \int_\mathcal{B} d^2 x \, \lambda_{o}  \ssg f(\sigma_{ab}, N,
   V^a)\, .  
\end{equation}
Readers familiar with \cite{BY}, will immediately recognize $E_{d/dt}$
with this $F(t)$ as the Brown-York energy, including the usual
reference term. In contrast, there is no freedom to add a reference
term to the angular momentum expression. Its most general form is
\be 
J'_\phi = \int_\mathcal{B} d^2 x \ssg \phi^a j_a \, ,
\ee
unchanged from equation (\ref{angmom}). In the Brown-York derivation
a freedom remains in this definition. 

Although specific choices of $F(t)$ will not be our major concern
here, a few words are in order.  The $F(t)$ is a term in the
expression for energy that cannot be determined by the
formalism. Indeed as far as our formalism is concerned all $F(t)$ are
equally valid. Thus, a particular $F(t)$ can be chosen to suit the
needs of a particular problem. Popular choices in the literature are
usually based on the requirement that $E_{d/dt}$ vanish for some
particular space-time (for example flat space or AdS). The reader is
directed to any of the papers listed in \cite{BYapp} for further
discussion of this point and other applications of this energy
expression.

\section{Comparison with action arguments}
\label{s4}

Let us now compare our calculations with those used in the Brown-York
derivation of the quasi-local energy. Their original paper \cite{BY}
pursued two lines of argument, both of which follow from an analysis
of the gravitational action $I$ and the boundary conditions which must
be imposed on quasi-local boundaries so that the action principal will
be well defined. From the action they proceeded in two directions to
obtain the quasi-local energy. The first used Hamilton-Jacobi type
arguments and followed a more qualitative path, while the second used
a Legendre transform to derive a Hamiltonian functional from the
action. In the next few paragraphs we will review the Legendre transform
route.

\begin{figure}
\center{\resizebox{!}{8cm}{\includegraphics{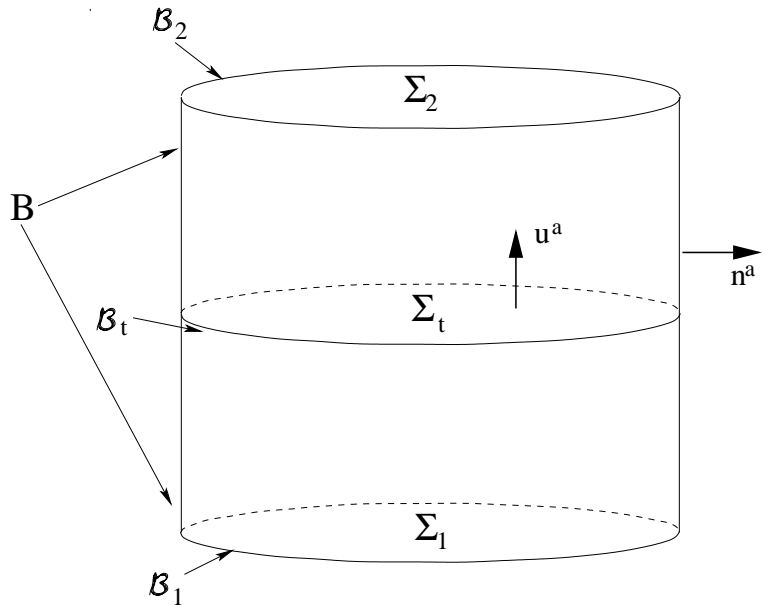}}}
%\vspace{8cm}
\caption{The space-time region $M$ with its boundaries, foliation, and
associated vector fields.}
\label{M}
\end{figure}

To begin, let us recast the three-dimensional constructions of the
previous section into four-dimensional space-time. We consider a region
of space-time $M$ that is bounded by a time-like boundary $B$ and two
space-like boundaries $\Sigma_1$ and $\Sigma_2$ as shown in figure
\ref{M}. The metric over $M$ will be $g_{ab}$ and the associated
covariant derivative is $\nabla_a$. The $\Sigma$ of the last section
is now replaced with space-like surfaces $\Sigma_t$ that foliate
$M$. We assume that $\Sigma_1$ and $\Sigma_2$ are leaves of that
foliation. The $\Sigma_t$ induce a foliation $\mathcal{B}_t$ of $B$,
and these $\mathcal{B}_t$ replace $\mathcal{B}$.  The boundary metrics
$h_{ab}$, $\gamma_{ab}$, and $\sigma_{ab}$ are all induced by $g_{ab}$
and may be calculated as
\bea
h_{ab} &=& g_{ab} + u_a u_b, \\
\gamma_{ab} &=& g_{ab} - n_a n_b \, , \quad \mbox{and} \\
\sigma_{ab} &=& g_{ab} + u_a u_b - n_a n_b,
\eea
where $n_a$ is the outward-pointing space-like unit normal to $B$ and
$u_a$ is the future pointing time-like unit normal to the $\Sigma_t$.
In defining these metrics we have built in the orthogonality
assumption that $u^a n_a = 0$.  Allowing non-orthogonal intersections
increases the complexity of the calculations but doesn't substantially
change the following results \cite{nopaper}.

To understand the Brown-York arguments, let us begin with an analysis
of the well-known trace-K action for the quasi-local region $M$:
\bea
\label{action}
I &=& \frac{1}{2 \kappa} \int_M d^4 x \sqrt{-g} \mathcal{R} 
+ \frac{1}{\kappa} \int_\Sigma d^3 x 
\sqrt{h} K - \frac{1}{\kappa} \int_B d^3 x \sqrt{- \gamma} \Theta \, .
\eea
In the above, $\mathcal{R}$ is the Ricci scalar corresponding to
$g_{ab}$.  Following the Brown-York sign convention for extrinsic
curvatures, $K = - \nabla_a u^a$ is the trace of the extrinsic
curvature of $\Sigma_{1,2}$ in $M$ while $\Theta = - \nabla_a n^a$ is
the trace of the extrinsic curvature of $B$ in $M$. Further
$\int_\Sigma = \int_{\Sigma_2} - \int_{\Sigma_1}$ and $\kappa = 8 \pi
G$.

The first variation of this action with respect to the metric $g_{ab}$
is (see any standard text on relativity, for example \cite{wald}),
\be
\label{firstVar}
\delta I = \frac{1}{2 \kappa} \int_M d^4 x \sqrt{-g} (G_{ab} + 
\Lambda g_{ab})  \delta g^{ab}  + \int_\Sigma d^3 x  \left( P^{ab}
\delta h_{ab} \right)  
+ \int_B d^3 x \left( \pi^{ab} \delta \gamma_{ab} 
\right). \nn
\ee
Note that in four-dimensions, $P^{ab}$ is not an independent field as
in the Hamiltonian formulation, but is instead 
\be
\label{Pab}
P^{ab} := \frac{\sqrt{h}}{2 \kappa} \left( K h^{ab} - K^{ab} \right),
\ee
where $K_{ab} = -{h_a}^c {h_b}^d \nabla_c u_d$ is the extrinsic
curvature of $\Sigma_{1,2}$ and $K$ is the trace of $K_{ab}$ as
discussed above.  Similarly,
\be 
\pi^{ab} := - \frac{\sqrt{-\gamma}}{2 \kappa} 
\left( \Theta \gamma^{ab} - \Theta^{ab} \right)
\ee
is an equivalent tensor density defined by the surface $B$, with
$\Theta_{ab} = -{\gamma_a}^c {\gamma_b}^d \nabla_c n_d$ being the
extrinsic curvature of $B$.

Then, if the boundary metrics $h_{ab}$ and $\gamma_{ab}$ are fixed
so that
\be
\delta \gamma_{ab} = 0  \quad \mbox{ and} \quad \delta h_{ab} = 0,
\ee
the variation of the action vanishes on-shell. Note however, that this
means that rotations or translations of the boundary are not allowed
variations of the action, unless those rotations/translations are
generated by Killing vector fields. We will come back to this point
momentarily, but for now we note that requiring that the variation of
the action vanish and boundary conditions be met gives the Einstein
equations in the standard way. Further note that with these boundary
conditions, free functionals of the boundary metrics may be added to
the action without affecting its first variation --- since those terms
are fixed, the variations of the functionals will vanish. As such, the
exact form of the action is ambiguous and
\be\label{actionfreedom}
I' = I + I_{o} [\gamma_{ab}, h_{ab}],
\ee
is an equally valid action for any functional $I_{o}$ of the boundary
metrics, in the sense that fixing the same boundary conditions and
demanding that $\delta I' = 0$ will also generate the usual equations
of motion.

Next, in preparation for applying a Legendre transform and so
obtaining a Hamiltonian, we take note of the foliation of $M$. This
gives an (imposed) notion of ``instants of simultaneity.''  A
time flow is put on $M$ in the guise of a vector field $\part^a$ that
satisfies $\part^a \partial_a t = 1$ and which lies in $TB$ everywhere
on $B$ --- that is, the time flow generates the boundary. From our
orthogonality assumption that $u^a n_a = 0$, we know that $u^a$ is
both the unit normal to the $\Sigma_t$ as well as to the
$\mathcal{B}_t$ in $B$. Thus, we may write
\be
T^a = N u^a + V^a \, ,
\ee
for some lapse $N$ and shift vector field $V^a$ that is everywhere an
element of $T \mathcal{B}_t$ over $B$.  Making use of this time
evolution, we can perform a Legendre transform on the action to
obtain: 
\bea
\label{MgravActdecomp}
I &=& \int dt \left\{ -H_t + \int_{\Sigma_t}  d^3 x 
\left( P^{ab} \Lie_\part h_{ab} \right) \right\} 
\eea  
where
\be
\label{gravHam}
H_t = 
\int_{\Sigma_t} d^3 x [N \cH + V^a \cH_{a}] 
+ \int_{\mathcal{B}_t} d^2 x \sqrt{\sigma} (N \varepsilon - V^a j_a).
\ee
Now, $\varepsilon$ retains its earlier value (see equation
(\ref{ejs})) but in the context of this four-dimensional approach
$j_a$ can be written as
\be
j_a = \frac{1}{8 \pi G} {\sigma_a}^{b} u^c \nabla_b n_c,
\ee
and so is proportional to the connection on the normal bundle to
$\mathcal{B}_t$. Referring back to (\ref{ejs}) it is probably also
worthwhile to note that in four-dimensions $a_a = u^b \nabla_b u_a$ is
the acceleration of the unit normal vector field $u_a$ along its
length.

The quasi-local gravitational Hamiltonian for the region of space-time
$M$ is given by $H_{t}$.  However, there is no reason that the
functional $I_{o}[\gamma_{ab}, h_{ab}]$ appearing in (\ref{actionfreedom})
should decompose into an integral with
respect to $t$ and so give us a Hamiltonian reference term that is 
local in time --- extra assumptions have to be made in order for this to 
be true.  The usual requirement is that $I_{o}$ be a linear functional
of the lapse and shift \cite{BY} which guarantees that the energy
density $\varepsilon$ and angular momentum density $j_{a}$ are
independent of lapse and shift.  Then functions of the form
\be
   H_{o} = \int_{\mathcal{B}} d^2x \left\{N f(\sigma_{ab}) + V^a
   f_a(\sigma_{ab})\right\}\,, 
\ee
where $f$ is a free function of $\sigma_{ab}$ and $f_a$ is a free
vector valued function of the same, may be added to the Hamiltonian.

Comparison of (\ref{et}) and (\ref{gravHam}) shows that the phase
space methods of section \ref{s3} and the action arguments given here
derive the same quasi-local energy, up to assumptions made about the
boundary conditions. Recall that in the phase space calculation, we
found free functionals of time only --- demanding a connection with
the boundary data allowed us to write them in the form of equation
(\ref{boundterm})
\be
F(t) = \int_{\mathcal{B}} d^2 x f[\sigma_{ab}, N, V^a] \nn.
\ee
By contrast, the Brown-York freedom is more general and allows free
functionals of the boundary metrics in the action without any need to
integrate.  It is only after extra assumptions that they can be broken
up into Hamiltonian reference terms that are independent of local
details of the geometry of $B$.  This difference in boundary terms
arises because the boundary conditions that we have imposed are not
the same. The Brown-York conditions are more restrictive than ours
(and so allow more freedom in the free functionals).

To obtain a fairer comparison, let us weaken the boundary conditions
in the action formulation so that they are equivalent to those used in
section \ref{s3}.
\begin{enumerate}

\item
Instead of rigidly fixing the metrics on the boundaries, we allow
variations which are foliation preserving translations/rotations of
the boundary. Thus, the overall geometry of the boundaries will be
fixed, although particular features of the geometry will not be fixed
to particular points of the manifold. To do this, we allow the
variations to act as diffeomorphisms on the boundary metrics, with the
restriction that they should map $B$ into itself. Then,
\be
\delta \gamma_{ab} = \Lie_Y \gamma_{ab} \quad \mbox{ and} \quad
\delta h_{ab} = \Lie_Z h_{ab} , 
\ee
for some vectors $Y^a \in TB$ and $Z^a \in T \Sigma$ for which $Y^a =
Z^a \in \mathcal{B}_{1,2}$ on those corner two-surfaces.

\item
At the same time, 
we impose the new condition that the diffeomorphism constraint should
hold on the boundary surfaces $\Sigma_1$, $\Sigma_2$, and $B$. That is
\be
D_a P^{ab} = 0 \quad \mbox{ and } \triangle_a \pi^{ab} = 0,
\ee
where, as before, $D_a$ and $\triangle_a$ are the induced covariant
derivatives on $\Sigma_{1,2}$ and $B$ respectively. Of course, these
constraints hold automatically on-shell. In the action formulation
however, variations are not restricted to being on-shell and we will need
to impose these conditions in the following calculation.
\end{enumerate}

\noindent Then, with these new conditions:
\bea
\delta I 
&=& 
\int_B d^{3}x \, \pi^{ab} \Lie_Y \gamma_{ab} 
+ \int_{\Sigma_2-\Sigma_1} d^{3}x \, P^{ab} \Lie_Z h_{ab}\\
&=& 2 \int_B d^{3}x \, \pi^{ab} \triangle_a Y_b  
+ \int_{\Sigma_2-\Sigma_1} d^{3}x \, P^{ab} D_a Z_b \nn  \\
&=& 2 \int_B d^{3}x \, \triangle_a ( \pi^{ab} Y_b )   
+ \int_{\Sigma_2-\Sigma_1} d^{3}x \, D_a (P^{ab} Z_b) \nn \\
&=& 2 \int_{\mathcal{B}_2 - \mathcal{B}_1} d^2 x \ssg
\left[ - \frac{1}{\sqrt{-\gamma}} u_a \pi^{ab} Y_b + \frac{1}{\sqrt{h}} n_a
P^{ab} Z_b \right] \nn \\
&=&  \frac{1}{\kappa} \left( \int_{\mathcal{B}_2 - \mathcal{B}_1} d^2 x \ssg
Y^b \left[ u^a \nabla_b n_a + n^a \nabla_b u_a \right] \right) \nn \\
&=& 0. \nn
\eea
The second line applies the representation of a Lie derivative in
terms of the covariant derivatives $\triangle_a$ and $D_a$ of $B$ and
$\Sigma_{1,2}$ respectively. The third uses the fact that the
diffeomorphism constraint holds on the boundary. The fourth integrates
bulk terms out to the boundaries of $B$ and $\Sigma_{1,2}$, while the
fifth uses the fact that $Y^a=Z^a \in T{\mathcal{B}_{1,2}}$ on those
boundaries. Thus, with these modified boundary conditions, the action
also vanishes on-shell.\footnote{The essentials of this proof that the
action is invariant with respect to diffeomorphisms that map the
boundary into itself may be found in \cite{BYLrecent}, where they
occur in the course of a demonstration that the Bianchi (and other)
identities may be derived from the action principal.}

Given these boundary conditions, there are other actions for which
$\delta I = 0$ and also give the same equations of motion. In
particular, consider any reference term which is of the form
\be
I_{o} = - \int^{t_2}_{t_1} dt F(t),
\ee
where $t$ is some labeling of the foliation which is equal to
$t_1$ on $\Sigma_1$ and $t_2$ on $\Sigma_2$. $F(t)$ is a functional
that depends on the foliation surface only. Then $\delta I_{o} =0$ for
the variations that we have considered. On applying the Legendre 
transform, we will find that for $I + I_{o}$ the Hamiltonian becomes
\bea
H_t = F(t) + 
\int_{\Sigma_t} d^{3}x \, [N \cH + V^a \cH_{a}] 
+ \int_{\mathcal{B}_t} d^2 x \sqrt{\sigma} (N \varepsilon - V^a j_a) \, ,
\eea
which is the same form that we found with our phase space
analysis. Thus, when we impose the same boundary conditions for the
two approaches, we obtain the same value for the Hamiltonian
functional --- as would be expected. As we have seen above, with the
standard Brown-York conditions which fix the metric at each point in
the boundary the free functionals can depend on specific local
features of the intrinsic geometry (equivalently, we could think of a
coordinate system imposed on the boundary with the functional being
allowed to depend on the coordinate system as well as the
metric). With the new boundary conditions however, that freedom has
changed and the free functionals can only depend on global geometric
features (for example integrals of geometric invariants over the
boundary).

\section{Discussion}
\label{s5}

In this paper we have introduced a phase space description of general
relativity in a manifold with boundary.  The boundary is equipped with
a preferred foliation and a metric which is fixed up to
diffeomorphisms which preserve this foliation. Therefore, we must
allow phase space variations which act as diffeomorphisms of the
boundary metric.  This can be accomplished by introducing a
coordinatization of the boundary and extending the phase space to
include both the time and spatial coordinates of the boundary and
their conjugate momenta.  Then, in the extended phase space, we are
able to discuss variations which induce diffeomorphisms of the
boundary, even if they are not symmetries of the boundary metric.  In
particular, time evolution is an allowed variation even if the
boundary metric is time dependent.  Similarly, rotation along a vector
field which is \textit{not} a Killing vector of the boundary metric is
also permitted.

Furthermore, we are able to associate ``conserved charges'' with these
motions.  In the case where the motion is not a symmetry of the
boundary, the ``conserved charge'' will not be constant.  Hence, the
conserved charges cannot be equal to the on-shell value of the
Hamiltonian (which is by definition a constant).  Instead, the on-shell 
values of the momenta conjugate to the coordinates give the
``conserved charges'' associated with these motions.   In particular,
if the variation induces a time translation at the boundary, the
corresponding conjugate momentum is equal to the energy of the
boundary.  Similarly if the variation produces a rotation of the
boundary, the conjugate momentum is the angular momentum of the
boundary.  These quantities need not be conserved: they can vary as
you move up the boundary.  In addition, the final expressions for the
energy and angular momentum are \textit{independent} of the choice of
coordinate system of the boundary, as one would hope.

The expressions for energy and angular momentum obtained from the
Hamiltonian formulation are the same as those obtained in the
Brown--York formalism (up to the free functionals where a difference
arises due to the slightly different boundary conditions).  This is
unsurprising since both methods describe the same situation, just in a
different language.  In particular, the energy obtained is not
uniquely fixed and there is the freedom to add to it a function of the
boundary data. When we harmonize boundary conditions as we did in the
last section, this is the same freedom as found in the Brown--York
formalism and can be fixed in the usual manner.  It should be possible
to extend the formalism presented here to the case of non-orthogonal
boundaries as considered in \cite{nopaper}, and work is already under
way to do so.
\footnote{In fact, in his alternative formulation, Kijowski \cite{kij}
has already shown that the non-orthogonal ``angle" parameter may be
included as a configuration variable in the phase space. }
Additionally, we have required throughout that
variations preserve the foliation of space-time.  It is likely that
this condition can also be weakened or removed entirely.

The approach we have described above should have several applications.
Indeed, it was initially motivated by a desire to extend the isolated
horizon framework \cite{ih} to include the physically interesting case
of dynamical horizons.  In order to describe these horizons within a
Hamiltonian framework, it is necessary to allow for a varying black
hole mass and angular momentum.  Making use of the extended phase
space methods introduced here will make this possible \cite{bf}.  A
second application would be to consider the Hamiltonian formulation of
general relativity with a boundary at null infinity.  Again, the mass
and angular momentum are not constant as they can be radiated away.
Therefore, the framework here would again be relevant.  It would be
interesting to recast the results of Wald and Zoupas \cite{wz} in
terms of the phase space description presented here.

\section*{Acknowledgments}

We are grateful to Abhay Ashtekar and Badri Krishnan for stimulating
discussions. Robert Mann read a preprint of this paper and supplied us
with useful suggestions on improving the presentation.  The comments
of the referees have also been very useful, especially in alerting us
to relevant papers in the literature with which we were not familiar.
We would also like to thank those who have attended seminars given by
IB on this topic. Audiences at the Perimeter Institute and Memorial
University have asked questions and made comments that have influenced
our thinking on some important points.  SF was supported by a Killam
postdoctoral fellowship at the University of Alberta. During the
initial stages of this work, IB was also at the University of Alberta
and supported by an NSERC PDF award. For the final stages, he
acknowledges the support of Memorial University of Newfoundland.

\appendix

\section{Constraints and time evolutions}
\label{appA}

For completeness we include the expressions for $[h_{ab}]_{(N,V)}$,
$[P^{ab}]_{(N,V)}$, $\mathcal{H}$, and $\mathcal{H}_b$ (in the absence
of matter and a cosmological constant). Keep in mind that the tensor
(and tensor density) fields in these expressions all live on $\Sigma$
and its assorted tensor bundles. Further, $D_a$ is the intrinsic
covariant derivative over $\Sigma$ and $\kappa = 8 \pi G$, where $G$
is the gravitational constant.
\label{messy}
\bea
\left[ h_{ab} \right]_{(N,V)} &\equiv& 
   \frac{4 \kappa N}{\sqrt{h}}[P_{ab} - \frac{1}{2} P h_{ab}]
   + \Lie_V h_{ab}, \label{dhdt} \\
\left[ P^{ab} \right]_{(N,V)} & \equiv & 
   - \frac{\sqrt{h}}{2 \kappa} \left( N\, {}^{(3)}G^{ab} 
   - \left[ D^a D^b N - h^{ab} D_c D^c N \right] \right) + \Lie_V
   P^{ab} \label{dPdt}\\ 
&& + \frac{N \kappa}{\sqrt{h}} \left(
[ P^{cd} P_{cd} - \frac{1}{2} P^2 ] h^{ab} 
- 4 [ P^{c(a}P_c^{\ b)} - \frac{1}{2} P P^{ab} ] \right), \nn \\
\cH &\equiv& -\frac{\sqrt{h}}{2 \kappa} 
R + \frac{2 \kappa}{\sqrt{h}}
\left(P^{a b} P_{a b} - \frac{1}{2} P^2 \right), \hspace{.5cm} \label{H} 
\mbox{ and} \\
\cH_b &\equiv&  -2 D_b {P_{a}}^{b}. \hspace{.5cm} \label{Ha}
\eea
In the above ${}^{(3)} G_{ab} = R_{ab} - \frac{1}{2} R g_{ab}$, where
$R_{ab}$ is the Ricci tensor on $\Sigma$ and $R$ is its contraction
with the metric $h_{ab}$.

Full derivations of the above results from the four-dimensional
geometry of solutions to the Einstein equations can be found in most
standard relativity texts, or using this notation in
\cite{mythesis}. Essentially though, they come from thinking of
$\Sigma$ as one leaf in the foliation of a solution of the Einstein
equations $(M,g_{ab})$ into space-like hypersurfaces (as discussed in
section \ref{s4}). If $u^a$ is the future pointing unit normal to
$\Sigma$, the time evolution given by $(N,V^a)$ corresponds to that
generated by the vector field $T^a = N u^a + V^a$. Further $P^{ab}$ is
no longer an independent variable, but instead is closely connected to
the extrinsic curvature of $\Sigma$ in $M$ (see equation (\ref{Pab})).

The Einstein equations tell us that in empty space, $G_{ab} = 0$.
Then, the Hamiltonian constraint (\ref{H}) is equivalent to the
statement $G_{ab} u^a u^b = 0$, the diffeomorphism constraint
(\ref{Ha}) is equivalent to $h_a^b G_{bc} u^c = 0$, and the evolution
equation for $P^{ab}$ (\ref{dPdt}) is equal to $\Lie_T P^{ab}$ with
$h_a^c h_b^d G_{cd} = 0$ being used to rewrite the result in terms of
quantities defined entirely on $\Sigma$. Finally, the evolution
equation for $h_{ab}$ is equivalent to $\Lie_T h_{ab}$.

\section{Variational calculation}
\label{appb}

In this appendix we show that if $\lambda_o$ is any function on
$\Sigma$ which is constant on $\mathcal{B}$ and $\lambda^a \in
T\Sigma$ is any vector field that lies in $T \mathcal{B}$ on
$\mathcal{B}$ so that
\be
  \delta_\Lambda P^{ab} \equiv \lambda_{o} [P^{ab}]_{(N,V)} +
  \Lie_{\lambda} P^{ab}  \quad \mbox{and} \quad 
  \delta_\Lambda h_{ab} \equiv \lambda_{o} [h_{ab}]_{(N,V)} +
  \Lie_{\lambda} h_{ab} \, , \label{hL} 
\ee
then
\bea \label{appbres}
&&\int_\Sigma d^{3}x \left[ (\delta_\Lambda h_{ab})(\delta P^{ab}) - 
   (\delta_\Lambda P_{ab})(\delta h_{ab}) \right] = \\
&&\quad \delta \left(\int_{\Sigma} d^{3}x \left[ \lambda_{o} N \,
   \cH + (\lambda^a + \lambda_{o} V^{a})\cH_{a} \right] 
   +  \int_{\mathcal{B}} d^{2}x \, \sqrt{\sigma} \left[\lambda_{o} 
   (N\varepsilon - V^{a} j_{a}) - \lambda^{a} 
   j_{a}\right] \right) \nonumber \\
&&\quad - \int_{\Sigma} d^{3}x \left[ \delta(\lambda_{o} N) \cH +
   \delta(\lambda^{a} + \lambda_{o} V^{a})\cH_{a} \right] \nonumber \\  
&&\quad - \int_\mathcal{B} d^{2}x \, \sqrt{\sigma} \left[
   \varepsilon \, \delta (\lambda_{o}N) - j_{a}\, \delta(\lambda_{o}
   V^a + \lambda^a) - \frac{\lambda_{o} N}{2} s^{ab} \delta \sigma_{ab}
\right], \nonumber 
\eea
where
\bea
\varepsilon &:=& k/(8 \pi G), \nonumber \\
j_a &:=& - 2 \sigma_{ac} P^{cd} n_d /\sqrt{h}, \quad \mbox{and} 
\nonumber \\
s^{ab} &:=& (1/8 \pi G) \left( k^{ab} - (k - n^c a_c) \sigma^{ab}
\right) \, . \nonumber
\eea
$k_{ab} = - {\sigma_a}^c {\sigma_b}^d D_c n_d$ is the extrinsic
curvature of $\mathcal{B}$ in $\Sigma$ (note the sign convention),
$k=\sigma^{ab} k_{ab}$, and $a_c = \frac{1}{N} D_c N$.

The calculations needed to obtain this result are quite lengthy, but
luckily the bulk of them may be found in previous papers. The key
result may be found in \cite{mythesis} or \cite{BYLrecent}, and says
that for a general variation $\delta$ (there is no restriction to it
being on-shell),
\bea
&& \delta \left(\int_\Sigma d^3 x [N \mathcal{H} + V^a \mathcal{H}_a] 
   \right) =  \label{deltaH1}\\ 
&&\qquad \int_\Sigma d^3 x \left\{ \mathcal{H} \delta N +
   \mathcal{H}_a \delta V^a + [h_{ab}]_{(N,V)} \delta P^{ab} -
[P^{ab}]_{(N,V)} \delta h_{ab} \right\} \nn \\
&&\qquad  - \int_{\mathcal{B}} d^2 x \left\{ N \delta(\sqrt{\sigma} 
   \varepsilon) - V^a \delta( \sqrt{\sigma} j_a) + \frac{1}{2}
   \sqrt{\sigma} N s^{ab} \delta \sigma_{ab} \right\}, \nonumber 
\eea
Now, a careful examination of the calculations found in the above
sources, shows that if we replace the $N$ and $V^a$ on the left hand
side of the equation with more general terms --- for example any
scalar field $\alpha$ over $\Sigma$ and any vector field $\beta^a$
over $\Sigma$ that is restricted to be parallel to $\mathcal{B}$ on
that boundary, then the expression changes to become
\bea
   \delta \left(\int_\Sigma d^3 x [\alpha \mathcal{H} + \beta^a
      \mathcal{H}_a]\right) &=& \int_\Sigma d^3 x 
      \left\{ \mathcal{H} \delta \alpha + \mathcal{H}_a \delta
      \beta^a\right\} \\ 
   &+& \int_\Sigma d^3 x\left\{ (\delta_\xi h_{ab}) (\delta P^{ab}) 
      - (\delta_\xi P^{ab})(\delta h_{ab}) \right\} \nonumber \\
   &-& \int_{\mathcal{B}} d^2 x \left\{\alpha\, \delta(\sqrt{\sigma} 
       \varepsilon) - \beta^a \,\delta( \sqrt{\sigma} j_a) 
      + \frac{1}{2}\sqrt{\sigma} \alpha s^{ab} \delta \sigma_{ab}
   \right\}, \nonumber 
\eea
where $\xi = (\alpha/N) \frac{d}{dt} + \Lie_{\beta - (\alpha/N) V}$
and $\delta_\xi P^{ab}$ and $\delta_\xi h_{ab}$ are defined in an
analogous way to $\delta_\Lambda P^{ab}$ and $\delta_\Lambda h_{ab}$
in equation (\ref{hL}). Therefore, in order to obtain the desired
result (\ref{appbres}), we must take
\[ \alpha = \lambda_{o}\, N \quad , \quad \beta^{a} =
\lambda^a + \lambda_{o} V^{a} \, , \]
and re-express the surface term as 
\bea
   && \int_{\mathcal{B}} d^2 x \left\{\alpha \, \delta(\sqrt{\sigma} 
      \varepsilon) - \beta^a \delta( \sqrt{\sigma} j_a) +
      \frac{1}{2}\sqrt{\sigma} \alpha s^{ab} \delta \sigma_{ab} \right\} 
      =  \nonumber \\
   && \qquad \delta \left(\int_{\mathcal{B}} d^{2}x \sqrt{\sigma}
      \left[\alpha \varepsilon - \beta^{a} j_{a})\right] \right)
      \nonumber  - \int_\mathcal{B} d^{2}x \sqrt{\sigma} \left[\varepsilon
      \delta (\alpha) - j_{a} \delta(\beta^a) - \frac{\alpha}{2} s^{ab}
      \delta \sigma_{ab} \right] \,, \nn 
\eea
to obtain (\ref{appbres}), the desired result.

\section{The diffeomorphism constraint on the boundary}
\label{calc}

\noindent 
In this appendix we consider a solution to the full four-dimensional Einstein
equations over a region $M$ like that discussed at the beginning of section
\ref{s4}. We show that if
\begin{enumerate}

   \item the diffeomorphism constraint holds on a time-like surface $B$
   that has intrinsic metric $\gamma_{ab}$,

   \item $B$ is foliatied with closed, space-like, two-surfaces
   $\mathcal{B}_t$ and there exists a vector field $\part^a$ such that
   $\part^a [dt]_a = 1$, and

   \item $X^a$ is a vector field taking the form $X^a = X_{o} \part^a +
   \hat{X}^a$ where $X_{o}$ is a function of $t$ alone, while $\hat{X}^a$
   is any non-singular vector field over $B$ that is everywhere
   parallel to the $\mathcal{B}_t$ surfaces, 

\end{enumerate}
then
\bea \label{appcres}
&& \int_{\mathcal{B}_t} d^2 x \left( \varepsilon \Lie_X N - j_a \Lie_X
   V^a - \frac{N}{2} s^{ab} \Lie_X \sigma_{ab} \right) = \nonumber \\ 
&&\qquad \int_{\mathcal{B}_t} d^{2}x \, X_{o} \,\Lie_\part \left(\ssg [N
   \varepsilon - V^a j_a] \right) - \int_{\mathcal{B}_t} d^{2}x \,
   \hat{X}^a \, \Lie_\part \left(\ssg j_a \right)\, .  
\eea
where $N$ and $V^a$ are the usual lapse and shift defined so that
if $u^a$ is the forward-pointing unit normal to the $\mathcal{B}_t$
(that is $\part^a u_a < 0$), then $\part^a = N u^a + V^a$.
$\sigma_{ab}$ is the induced two-metric on the $\mathcal{B}_t$
surfaces, and $\varepsilon$, $j_{a}$ and $s_{ab}$ are defined in
(\ref{ejs}).  

We begin by introducing the conjugate variable to $\gamma_{ab}$,
namely $\pi^{ab}$ which is given by:
\be\label{pi}
\pi^{ab} = \frac{\sqrt{-\gamma}}{16 \pi G} \left(\Theta^{ab} - \Theta
\gamma^{ab}\right) \, ,
\ee
where $\Theta_{ab} = {\gamma_a}^c {\gamma_b}^d \nabla_c n_d$ is the
extrinsic curvature of $B$ with respect to the unit normal $n^a$.
Then, we can write $\varepsilon$, $j_{a}$ and $s_{ab}$ in terms of
$\pi^{ab}$ so that
\begin{eqnarray}
   \varepsilon &=& - 2 \pi^{ab} u_a u_b /\sqrt{-\gamma}\, , \nonumber \\ 
   j_a &=& 2 \sigma_{ab} \pi^{bc} u_{c} /\sqrt{-\gamma}\, , \,\,
      \mbox{and} \nonumber \\ 
   s_{ab} &=& - 2 {\sigma_a}^c {\sigma_b}^d \pi^{cd} /\sqrt{-\gamma} \,
      . \nonumber 
\end{eqnarray}

Now, taking $\triangle_a$ as the covariant derivative on $B$ we first
note that the diffeomorphism constraint implies that
\[ \triangle_a \pi^{ab} = 0\, , \]
and therefore
\be\label{base}
\int_{\mathcal{B}_t} d^{2}x \, \triangle_a ( \pi^{ab} X_b ) = 
\frac{1}{2} \int_{\mathcal{B}_t} d^{2}x \, \pi^{ab} \,\Lie_X
\gamma_{ab} \, .
\ee
The result (\ref{appcres}) arises when we break up each side of the
expression (\ref{base}) into parts parallel and perpendicular to the
foliation. To that end note that 
\[ \gamma_{ab} = \sigma_{ab} - u_a u_b \quad \mbox{and} \quad
\sqrt{-\gamma} = N \sqrt{\sigma} \, .\] 

Then, working on the left hand side we can show that
\bea
&& \sqrt{-\gamma} \triangle_a (\pi^{ab} X_b /\sqrt{-\gamma}) = -
   \Lie_\part ( \ssg u_a \pi^{ab} X_b /\sqrt{-\gamma} )  \nonumber \\
&&\qquad\qquad + \Lie_V (\ssg u_a \pi^{ab} X_b /\sqrt{-\gamma} ) +
   \ssg \, d_a (N {\sigma^a}_b \pi^{bc} X_c /\sqrt{-\gamma} )\, .  
\eea
Since the surfaces $\mathcal{B}_t$ are closed, when we integrate this
expression only the first term on the right hand side survives, giving
\be 
   \int_{\mathcal{B}_t} d^{2}x \, \triangle_a ( \pi^{ab} X_b ) = 
   \frac{1}{2} \int_{\mathcal{B}_t} d^{2}x \,\Lie_\part  \left( \ssg
   (X_{o} \left[ N \varepsilon -  V^{a} j_{a} \right] - \hat{X}^a j_a)
   \right) \, . 
   \label{lhs}  
\ee

Let us now turn our attention to the right-hand side of equation
(\ref{base}). We begin by recalling that $u_a = -N [dt]_a$.
Therefore, with $X_o$ constant on each slice and $\Lie_X (dt_a) =
(\Lie_T X^0) dt_a$ it is simple to show that
\be
\Lie_X u_a = \left(\frac{\Lie_X N}{N} + \Lie_\part X_{o} \right) u_a \, .
\ee
In particular, it then follows immediately that $\sigma^{ab}\Lie_X u_a
= 0$. Finally, by decomposing the metric $\gamma_{ab}$ in terms of
$u_{a}$ and $\sigma_{ab}$ we obtain:
\bea
   \pi^{ab} \Lie_X \gamma_{ab} &=& \sqrt{-\gamma} \left\{\varepsilon
      \, u^a \, \Lie_X u_a + j_a (\Lie_u X^a + \sigma^{ab} \Lie_X u_b)
      + \frac{1}{2} s^{ab} \Lie_X \sigma_{ab} \right\} \nn \\ 
   &=&  \ssg \left( \varepsilon \Lie_X N - j_a \Lie_X V^a -
      \frac{N}{2} s^{ab} \Lie_X \sigma_{ab} \right) \label{rhs} - \ssg
      j_a \Lie_\part \hat{X}^a \nonumber \\
   &&\qquad + \ssg (N \varepsilon - V^a j_a) \Lie_\part X_{o} \, .  
\eea
Equations (\ref{base}), (\ref{lhs}), and (\ref{rhs}) can then be
combined to give the promised result:
\bea
&& \int_{\mathcal{B}_t} d^2 x \left( \varepsilon \Lie_X N - j_a \Lie_X
   V^a - \frac{N}{2} s^{ab} \Lie_X \sigma_{ab} \right) = \nonumber \\ 
&&\qquad \int_{\mathcal{B}_t} d^{2}x \, X_{o} \,\Lie_\part \left(\ssg [N
   \varepsilon - V^a j_a] \right) - \int_{\mathcal{B}_t} d^{2}x \,
   \hat{X}^a \, \Lie_\part \left(\ssg j_a \right)\, .  
\eea

\end{document}